\documentclass[reprint,superscriptaddress,showkeys,prl]{revtex4-1}

\usepackage{color,soul}
\usepackage[margin=5pt,captionskip=-10pt,justification=RaggedRight,singlelinecheck=false]{subfig}
\usepackage{graphicx}
\usepackage{amsmath}
\usepackage{units}
\usepackage[version=3]{mhchem}
\usepackage{overpic}

\renewcommand\hl[1]{#1}

\begin{document}

\title{Theory of Nucleation in Phase-separating Nanoparticles}
\author{Daniel A. Cogswell}
\affiliation{Samsung Advanced Institute of Technology America, Cambridge, MA 02142, USA}
\affiliation{Department of Chemical Engineering, Massachusetts Institute of Technology, Cambridge, Massachusetts 02139, USA}
\author{Martin Z. Bazant}
\affiliation{Department of Chemical Engineering, Massachusetts Institute of Technology, Cambridge, Massachusetts 02139, USA}
\affiliation{Department of Mathematics, Massachusetts Institute of Technology, Cambridge, Massachusetts 02139, USA}
\date{\today}
\keywords{core-shell nanoparticles, nucleation, coherency strain, phase-field modeling, electrochemistry, \ce{LiFePO4}}

\begin{abstract}
The basic physics of nucleation in solid \hl{single-crystal} nanoparticles is revealed by a phase-field theory that includes surface energy, chemical reactions and coherency strain. In contrast to binary fluids, which form arbitrary contact angles at surfaces, complete ``wetting" by one phase is favored at binary solid surfaces. Nucleation occurs when surface wetting becomes unstable, as the chemical energy gain (scaling with area) overcomes the elastic energy penalty (scaling with volume).  The nucleation barrier thus decreases with the area-to-volume ratio and vanishes below a critical size, and nanoparticles tend to transform in order of increasing size, leaving the smallest particles homogeneous (in the phase of lowest surface energy).  The model is used to simulate phase separation in realistic nanoparticle geometries for \ce{Li_XFePO4}, a popular cathode material for Li-ion batteries, and collapses  disparate experimental data for the nucleation barrier, with no adjustable parameters. Beyond energy storage, the theory generally shows how to tailor the elastic and surface properties of a solid nanostructure to achieve desired phase behavior. 
\end{abstract}

\maketitle

Despite the widespread use of phase-separating nanoparticles for catalysis, ``smart materials", and energy storage~\cite{dresselhaus1981,Ferrando2008,bruce2008}, their complex phase behavior is just beginning to be understood~\cite{shirinyan2004,tang2010,Bazant2013}.  The most basic open question involves nucleation, which is difficult to observe experimentally and beyond the reach of {\it ab initio} molecular simulations. (A 20 nm nanoparticle has $\sim50,000$ atoms.) It is known that surfaces are important \cite{Ferrando2008}, but their precise role is unclear. In bimetallic nanoparticles, where phase transitions are triggered by changes in temperature~\cite{Lee2004}, anomalous melting point depression in silica-gold core-shell nanoparticles persists to unexpectedly large particle sizes \cite{Dick2002}. Enhanced interdiffusion in gold-silver core-shell nanoparticles cannot be explained by size-dependent melting point depression \cite{Shibata2002}. 

For Li-ion battery nanoparticles, where phase transitions occur by ion intercalation, the literature is full of contradictions. The canonical phase-separating cathode material is \ce{LiFePO4}, which exhibits low power in micron-sized particles \cite{Padhi1997} but can achieve very high rates in nanoparticles  \cite{Wang2011}. Experimental measurements of the critical overpotential to initiate lithiation vary widely from  \unit[2]{mV} to \unit[37]{mV} \cite{Meethong2007, Meethong2007a,Yu2007,Lee2009, Dreyer2010, Come2011, Zhu2011,Safronov2011, Safronov2012,Farkhondeh2012}. Size dependence has also been reported \cite{Meethong2007,Farkhondeh2012}. Some experiments observe electrical signatures of nucleation and 1D growth above a critical particle size ~\cite{oyama2012}, while others attribute voltage hysteresis to mosaic phase separation among homogeneous particles without nucleation ~\cite{Dreyer2010}. Some theoretical studies suggest that nucleation at surfaces leads to ``intercalation waves" (moving phase boundaries) at low current~\cite{singh2008,Bai2011,Tang2011}, while others describe a ``solid-solution pathway'' without the possibility of nucleation~\cite{Malik2011}.   

In this article, we resolve these discrepancies by showing that nucleation in \hl{single-crystal} nanoparticles is size-dependent, occurring as a result of surface adsorption that leads to coherency strain (a long-range force). This mechanism implies a nucleation barrier that decreases linearly with the area-to-volume ratio, in quantitative agreement with a wide range of experimental data. We show that the data are consistent with both phase-field \cite{Cogswell2012} and \textit{ab initio} calculations \cite{Malik2011} that estimate a zero-current overpotential of about \unit[35]{mV} in bulk \ce{LiFePO4}.  \hl{The analysis presented in Methods is very general and could be applied to other multiphase nanostructures.}

\begin{figure}[t]
 \subfloat[]{\includegraphics[height=1.25in]{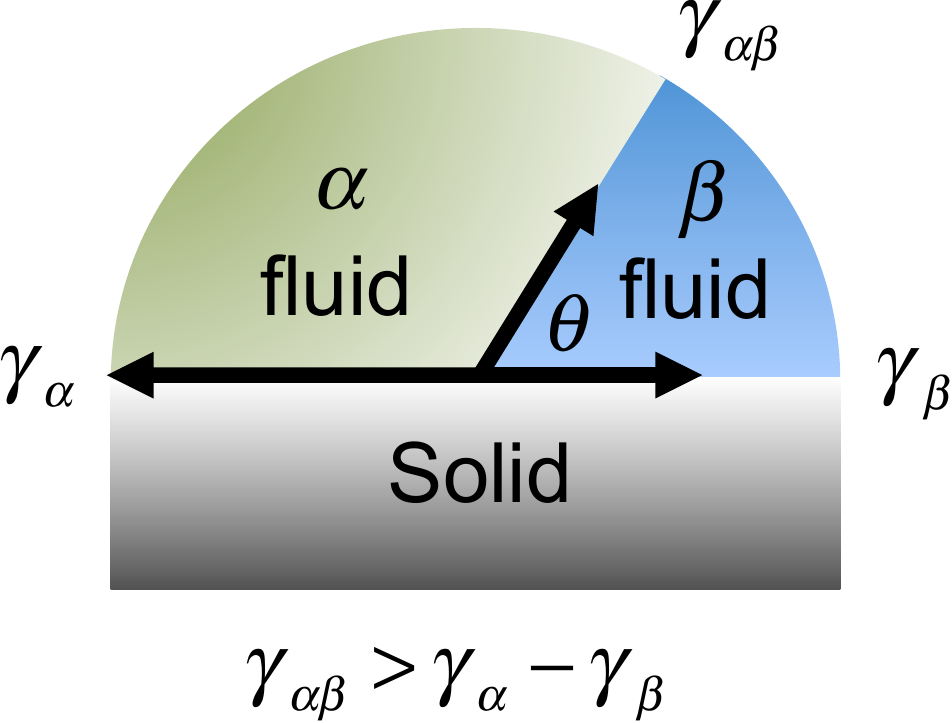}}
 \hspace{.05in}
 \subfloat[]{\includegraphics[height=1.25in]{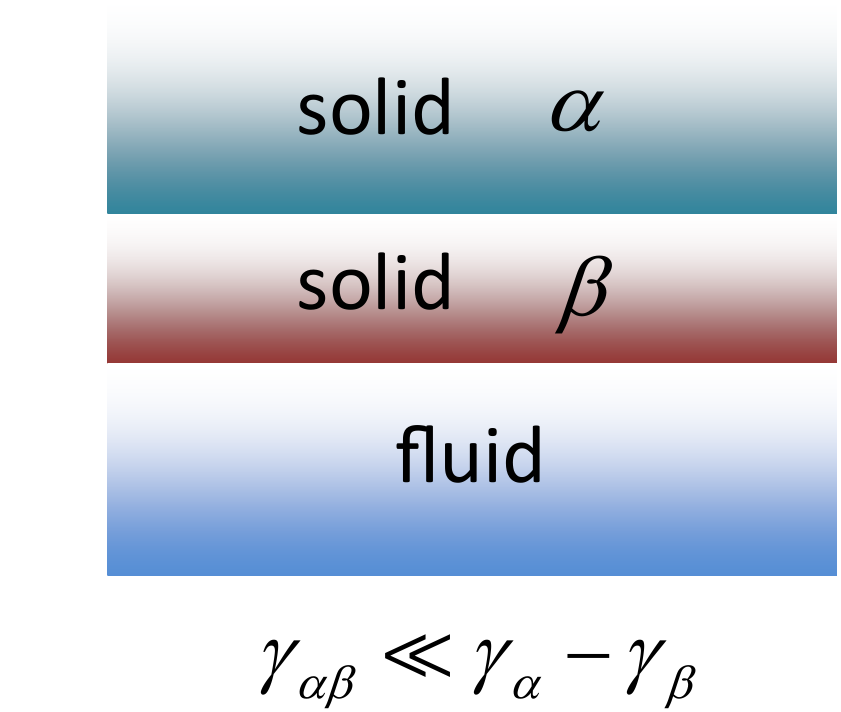}}
 \vspace{0.1in}
 \begin{tabular}{cccccc}
  \hline
  Orientation & (100) & (010) & (101) & (011) & (201)\\
  \hline
  $\gamma_\alpha-\gamma_\beta$ (mJ/m$^2$) & 260 & -400 & \hl{-}90 & 240 & 250\\
  $\theta (^\circ)$ & 0 & 180 & \hl{180} & 0 & 0\\
  \hline
 \end{tabular}
 \caption{ (a) Partial wetting by fluid $\beta$ displacing fluid $\alpha$ at a solid surface with contact angle $\theta$. (b) Complete ``wetting" by solid $\beta$ displacing solid $\alpha$ at a fluid surface. Table:
 Surface energies of \ce{FePO4} ($\alpha$) vs. \ce{LiFePO4} ($\beta$) calculated from first principles \cite{Wang2007}.  The smaller phase boundary energy~\cite{Cogswell2012}, $\gamma_{\alpha\beta}=\unit[39]{mJ/m^2}$, implies complete wetting by $\alpha$ or $\beta$ at each facet. }
 \label{fig:wetting}
\end{figure}

\begin{figure*}[ht]
\begin{center}
\subfloat[]{\includegraphics[width=.325\textwidth]{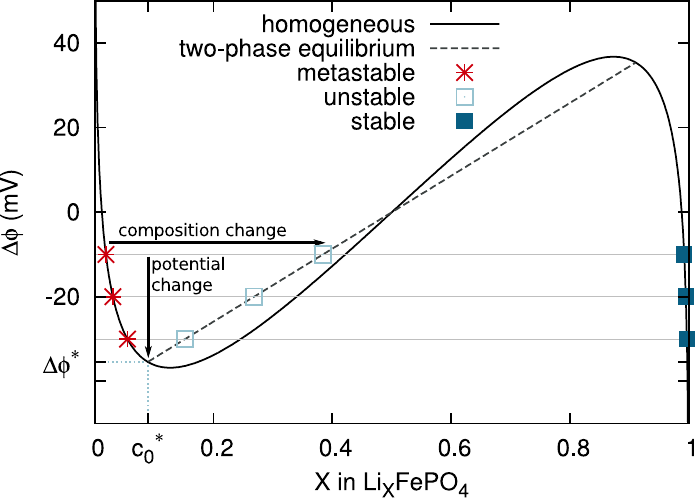}}
\subfloat[]{\includegraphics[width=.3\textwidth]{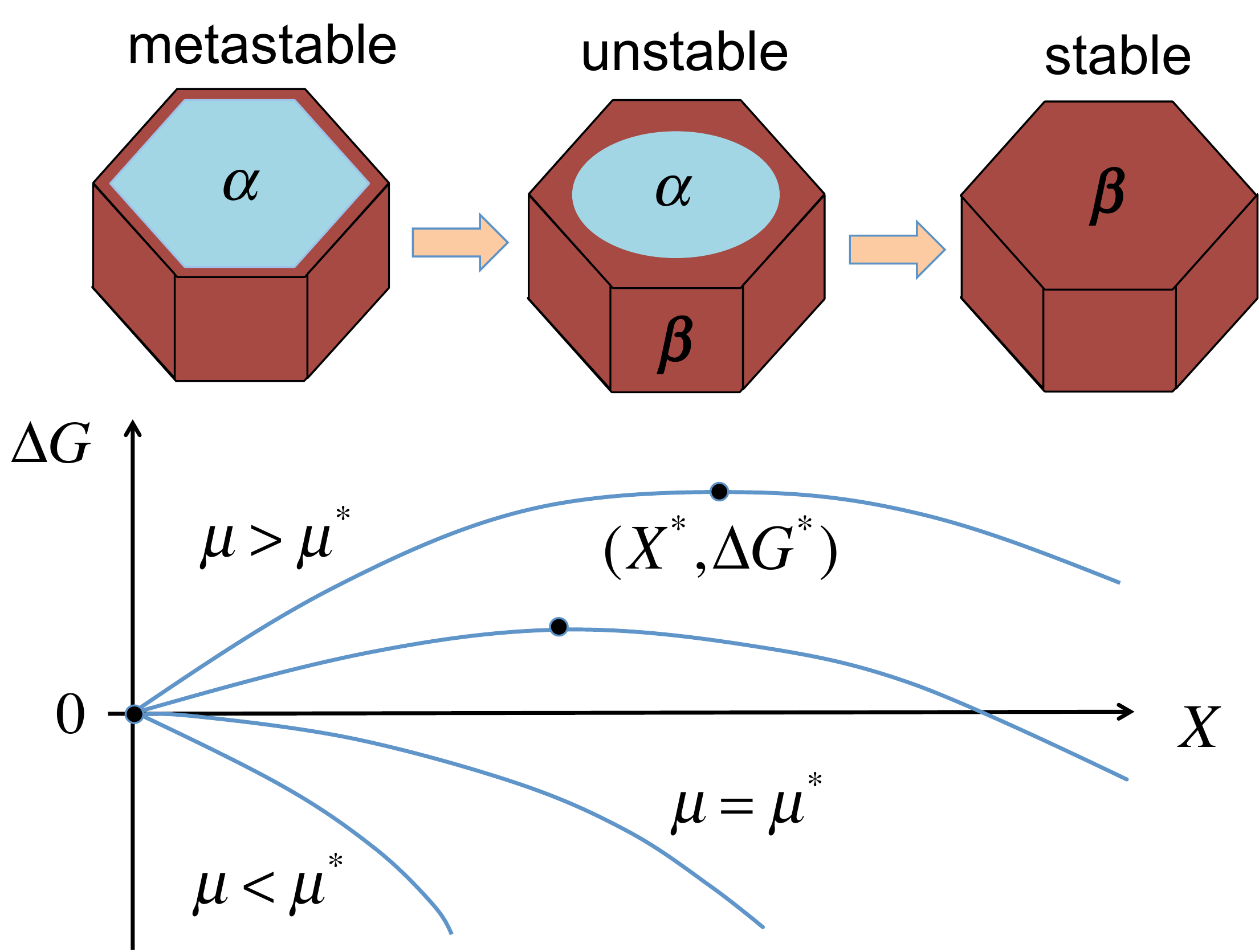}}
\subfloat[]{\includegraphics[width=.335\textwidth]{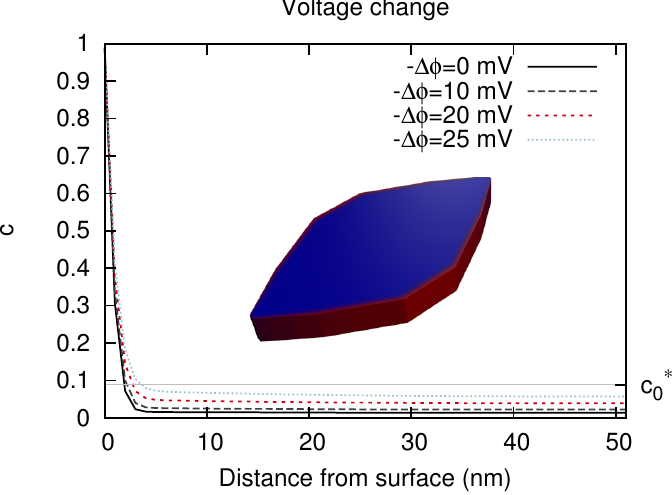}}
 \caption{(a) Equilibrium voltage curve for a binary intercalation particle, calculated with our phase-field model for \ce{LiFePO4}.  
For $\Delta\phi>\Delta\phi^*$ (or $\Delta\mu > \Delta\mu^*$), transformation from metastable to stable homogeneous states is nucleated by \hl{changes} in potential, composition, or temperature leading to unstable two-phase coexistence at the barrier.
(b) Sketch of nucleation \hl{at wetted surfaces due to changes in composition.} (c) Calculation of the surface adsorption profile in a \ce{LiFePO4} nanoparticle (C3 shape~\cite{Smith2012} with wetted side facets; inset) by voltage fluctuations, which increase the bulk concentration\hl{, $c_0(\Delta\phi)$,} until the surface layer becomes unstable at the coherent solubility limit \hl{$c_0^*$}.
  \label{Fig:potential}
 }
 \end{center}
 \end{figure*}

{\it Surface wetting in binary solids. --}
The wetting of a solid surface by binary fluids is one of the most studied problems in fluid mechanics \cite{Cahn1977,Bonn2001,Bico2002}, but a theory of surface ``wetting" in binary solids, which have coherency strain, has not been developed. Wagemaker \textit{et al.} \cite{Wagemaker2009, VanderVen2009a} modeled the effect of surfaces on the voltage curves and solubility limits of \ce{LiFePO4}, but assumed that phase boundaries form a contact angle with exterior surfaces and neglected coherency strain. Tang and Karma studied coherent spinodal decomposition at solid surfaces \cite{Tang2012}, but did not consider finite systems or dependence of surface energy on composition. Bai \text{et al.} simulated nucleation by surface wetting \cite{Bai2011}, but neglected coherency strain. 

Young's equation,
$\gamma_\alpha=\gamma_\beta+\gamma_{\alpha\beta}\cos\theta$, relates surface and interfacial tensions to the contact angle $\theta$ at a triple junction (Fig. \ref{fig:wetting}(a)). Because fluid-fluid and fluid-solid interfacial energies have the same order of magnitude, all contact angles are possible in binary fluids \cite{Bonn2001,Bico2002}, including complete wetting and de-wetting ($\theta=0^\circ,180^\circ$).

The surface of a binary intercalation compound involves equilibrium between two solid phases and a fluid (the electrolyte in battery). A stable triple junction ($0^\circ < \theta < 180^\circ$) is unlikely to form between two solid phases because the excess energy of the free surface (from broken bonds) is much larger than that of a coherent solid-solid interface (from stretched bonds) (Fig. \ref{fig:wetting}(b)).  A rule of thumb is that coherent interfaces have  $\gamma_{\alpha\beta} < \unit[200]{mJ/m^2}$, while solid surfaces have $\gamma_\alpha, \gamma_\beta >\unit[1]{J/m^2}$ \cite{Porter2009}.  The change in surface energy with composition thus dominates, and one solid phase will tend to completely wet the surface. If $\gamma'(c)=0$, a $90^\circ$ contact angle will form and coherent surface spinodal states exist \cite{Tang2012}, but this is not the typical situation.

This prediction is supported by the ``core-shell'' structures commonly observed in bimetallic nanoparticles with the low-$\gamma$ phase as the shell \cite{Ferrando2008}, as well as by first principles calculations of battery nanoparticles. The change in surface energy by between  $\alpha=$ \ce{FePO4}  and $\beta=$ \ce{LiFePO4} for common facets of the Wulff shape (Table \ref{fig:wetting}) greatly exceeds the phase boundary energy, $\gamma_{\alpha\beta}=\unit[39]{mJ/m^2}$, inferred from the thickness of striped phases~\cite{Cogswell2012}. As a result, each crystal facet tends to be fully lithiated ($\theta=0$) or delithiated ($\theta=180^\circ$).

\begin{figure*}[t]
\captionsetup[subfigure]{font=large,margin=0pt,captionskip=0pt,labelformat=empty,justification=centering,singlelinecheck=false}
\subfloat{\includegraphics[width=.1\textwidth]{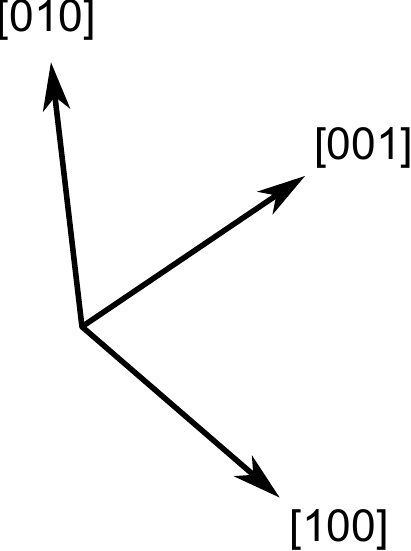}}
\subfloat[$X=0.1$]{\includegraphics[width=.18\textwidth]{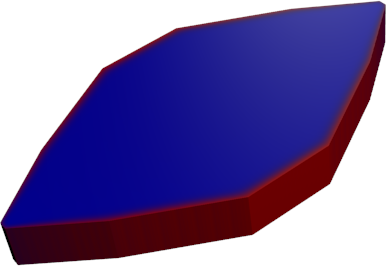}}
\subfloat[$X=0.25$]{\includegraphics[width=.18\textwidth]{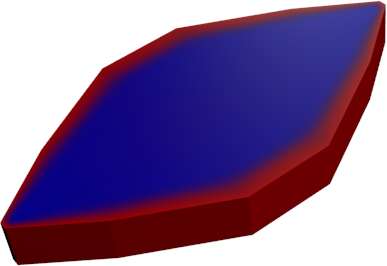}}
\subfloat[$X=0.5$]{\includegraphics[width=.18\textwidth]{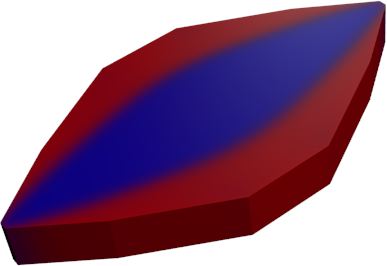}}
\subfloat[$X=0.75$]{\includegraphics[width=.18\textwidth]{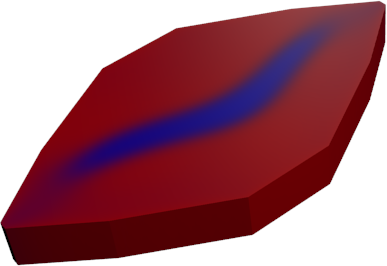}}
\subfloat[$X=0.99$]{\includegraphics[width=.18\textwidth]{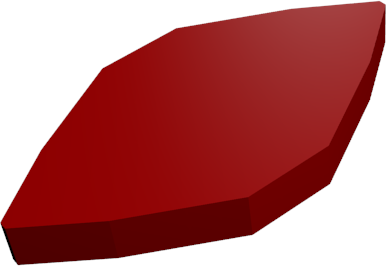}}
 \caption{ Phase-field simulation of galvanostatic lithiation of a Li$_X$FePO$_4$ nanoparticle (C3 shape \cite{Smith2012} with a 150 nm $\times$ 76 nm (010) facet (top)) illustrating nucleation \hl{at the coherent solubility limit ($X=0.09$), followed by growth from the lithiated side facets with interface alignment along (101) planes due to coherency strain.}  (Movie online)}
 \label{Fig:simulation}
\end{figure*}

{\it Nucleation at Solid Surfaces. --}  Consider a nanoparticle of volume $V$ in bulk phase $\alpha$ with a surface area $A$ wetted by molecules of phase $\beta$, in metastable equilibrium with diffusional chemical potential $\Delta\mu=\mu_\beta - \mu_\alpha$, where $\Delta\mu=0$ corresponds to chemical (or mosaic) two-phase coexistence. In the case of lithium ion intercalation in batteries~\cite{Bazant2013,Bai2011,Cogswell2012},  $\Delta\phi=\Delta\mu/e = \phi_e - \phi$ is the interfacial voltage (of electrons $\phi_e$ relative to ions $\phi$), and $\Delta\phi=0$ corresponds to mosaic phase separation across a set of homogeneous particles~\cite{Dreyer2010,ferguson2012}.  For a neutral species, $\Delta\mu$ can be controlled by adjusting the reservoir concentration of that species.

The transformation from the metastable phase $\alpha$ to the stable phase $\beta$ can be nucleated by fluctuations in composition, potential or temperature (Fig.~\ref{Fig:potential}(a)).  The nucleation barrier corresponds to the onset of (unstable) two-phase coexistence. At a critical potential,  $\Delta\mu^* = e \Delta \phi^*$, corresponding to the coherent solubility limit, the barrier for composition fluctuations vanishes ($\Delta G^b=0$), and the critical state is an unstable surface layer ($X_\beta=0$).

\begin{figure*}[t]
\subfloat[]{\includegraphics[width=.315\textwidth]{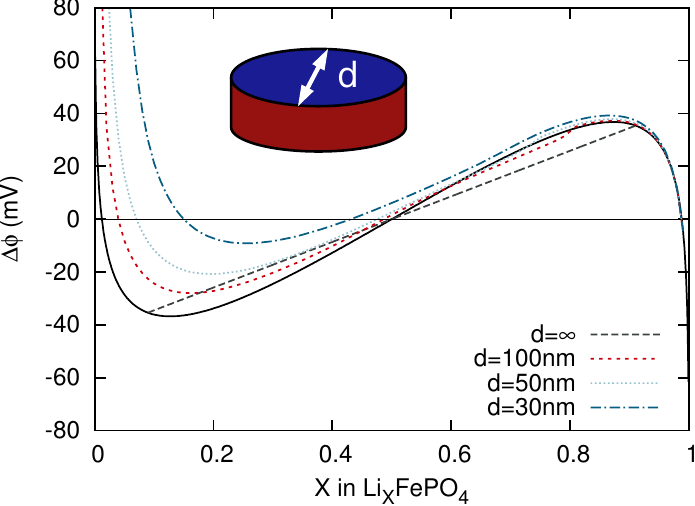}}
\hspace{.01\textwidth}
\subfloat[]{\includegraphics[width=.335\textwidth]{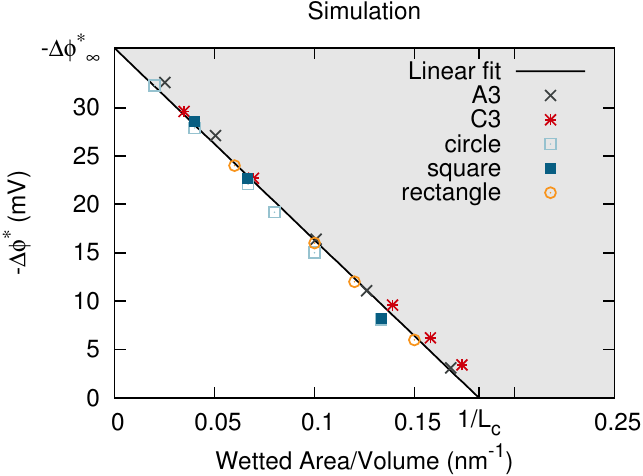}}
\subfloat[]{\includegraphics[width=.33\textwidth]{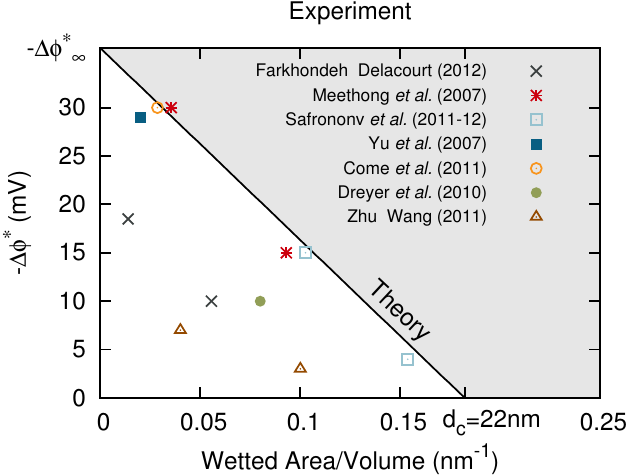}}
 \caption{Size dependence of nucleation in \ce{LiFePO4} nanoparticles.  
(a) Calculated discharge curves of circular cylinders of various diameters. (b) Calculated critical potential versus wetted area-to-volume ratio for different particle shapes, collapsing onto a master line. (c) Measured critical potentials versus reported particle sizes, falling on or below the theoretical limit for a perfect crystal (outside the gray region).}
 \label{Fig:critical_potential}
\end{figure*}

With battery nanoparticles, it is straightforward to control the voltage and observe a sudden current associated with \hl{phase transformation}. In this case, the bulk solid remains homogeneous at \hl{the bulk concentration $c_0(\Delta\phi)$} until the coherent solubility limit is reached at the critical potential $\Delta\phi^*$ (Fig.~\ref{Fig:potential}(c)). The Gibbs free energy 
\begin{equation}
G(c_0) = \left( f(c_0) + c_0 \Delta\mu \right) \rho_s V + \sigma(c_0) A
\label{Eq:free_energy}
\end{equation}
has contributions from the homogeneous Helmholtz free energy per volume, $\rho_s f(c_0)$ (where $\rho_s$ is the density of intercalation sites),  and the excess surface free energy of the wetted surface layer, $\sigma(c_0)$.
Setting $G^\prime(c_0^*)=0$ at the critical concentration $c_0^*=c_0(\Delta\phi^*)$, we obtain the critical potential,
\begin{equation}
-\Delta\mu^* = f^\prime(c_0^*) + \frac{\sigma^\prime(c_0^*)}{\rho_s} \frac{A}{V}
= -\Delta\mu^*_\infty\left( 1 -  \frac{L^*}{L} \right)
 \label{Eq:critical_potential}
\end{equation}
The critical potential in an infinite particle, $\Delta\mu^*_\infty=-f^\prime(c_0^*)$, is the difference between coherent and chemical (or mosaic) solubility due to the elastic strain exerted by the wetted surface on the bulk solid, which scales with volume $V$, as shown in Methods. Since surface energy scales with the wetted area $A$, the critical potential \hl{decreases with particle size, $L = V/A$, and vanishes below a critical value, $L_c=\frac{\sigma'(c_0^*)}{\rho_sf'(c_0^*)}$, which corresponds to $\Delta\mu^*=0$ (Fig.~{\ref{Fig:critical_potential}}). Below $L_c$ there is no barrier for transformation.}

{\it Phase-field model. --}   \hl{To more precisely determine $\Delta\phi^*_\infty$ and $L_c$}, we use the phase-field method, extended to include electrochemistry \cite{Bazant2013,Bai2011}, coherency strain \cite{Cogswell2012} and external surfaces \cite{Cahn1977}. Phase-field models accurately describe both homogeneous and heterogeneous nucleation \cite{Cahn1959,Granasy2007}.  For an inhomogeneous \hl{single-crystal} nanoparticle, the free energy is a functional of the intercalated ion concentration $c(\vec{x},t)$ and elastic displacement field $\vec{u}(\vec{x},t)$,
\begin{eqnarray}
G[c,\vec{u}] &=&   \int_V \left(f(c)+e\Delta\phi c\right) \rho_s dV + \int_A\gamma(c)\,dA    \nonumber \\
 & &  \int_V\left[ \frac{1}{2}\kappa(\nabla c)^2 +\frac{1}{2}C_{ijkl}\epsilon_{ij}\epsilon_{kl} \right]\, dV 
 \label{Eq:energy_functional}   
 \end{eqnarray}
where $\gamma(c)$ is the surface energy as a function of surface concentration, $\kappa$  the gradient energy coefficient,  $\epsilon_{ij}=\frac{\partial u_i}{\partial u_j}$  the strain tensor, and $C_{ijkl}$ the elastic stiffness tensor. Overpotential is defined as $\eta = \frac{\delta G}{e \rho_s \delta c}$ by the phase-field theory of electrochemical kinetics~\cite{Cogswell2012,Bazant2013}. Dynamical equations are given in Methods, and at equilibrium $G$ is minimized with $\eta=0$. 

Here we focus on the effect of surface energy, described by the natural boundary condition,
\begin{equation}
 \gamma'(c)=\hat{n}\cdot\frac{\delta G}{\delta \nabla c}=\hat{n}\cdot\kappa \nabla c
 \label{Eq:boundary_condition}
\end{equation}
which is necessary to enforce $\delta G=0$ \cite{Cahn1977,Granasy2007}. For the case of  a binary fluid in contact with a solid,  Cahn~\cite{Cahn1977} showed that  Eq. \ref{Eq:boundary_condition} is a generalization of Young's Law. For a typical binary solid (Fig.~\ref{fig:wetting}), \hl{the left hand side of Eq. {\ref{Eq:boundary_condition}} dominates, leading to complete wetting or dewetting of each facet, depending on the sign of $\gamma'(c)$. This can be seen by substituting Eq. {\ref{Eq:grad_c}} and performing the graphical analysis of Cahn {\cite{Cahn1977}}.}

The properties of the critical point are derived in Methods by defining $\sigma$ as the excess surface free energy of the adsorption layer (Fig. \ref{Fig:potential}(c)) and analyzing the equilibrium conditions $\frac{\delta G}{\delta c}=0$ and $\frac{\partial G}{\partial c_0}=0$. The derivation shows that $\sigma$ is independent of system system size, leading to Eq. \ref{Eq:critical_potential} for the critical potential. $\frac{d\sigma}{dc_0}$ is approximated as
\begin{equation}
\frac{d\sigma}{dc_0}(c_0^*)= \sqrt{\kappa B_0(c_0^*-X)}   \label{eq:slope}
\end{equation}
where $B_0$ is the elastic contribution to interfacial energy (scaling with $C_{ij}$), $c_0^*$ is the concentration at the coherent solubility limit, and $X=\frac{1}{V}\int_Vc\,dV$ \hl{is the state of charge of {\ce{Li_XFePO4}}}. For a very large system, a small amount of surface adsorption will have a negligible effect on $X$.  As the particle size decreases,   however, surface adsorption significantly changes $X$, which in turn creates an increased energy contribution from coherency strain.

Equation \ref{Eq:free_energy} is reminiscent of the free energy of forming a nucleus in classical nucleation theory (CNT), $\Delta G=V\Delta G_V+\gamma A$, but fundamentally different.  CNT is an approximation applied to a growing nucleus of a few nanometers, while Eq. \ref{Eq:critical_potential} is rigorous and applies to much larger nanoparticles with fixed size.  The intercalation analogue of the classical critical radius is \hl{a binary intercalation particle at the critical size $L_c$ where $\Delta\phi^*=0$.  Bulk free energy will be dominant in particles larger than $L_c$, but the phase state of particle smaller than $L_c$ will be controlled by surface energy.} Below the critical size, the nanoparticle will exist in the homogeneous state that minimizes surface energy.

\hl{ Since it is based on non-equilibrium thermodynamics~\cite{Bazant2013}, our model can be used to predict the dynamics of nucleation and growth, driven by surface reactions.  Phase transformation at constant current with a fixed transformation time (C rate) is illustrated in Fig. ~\ref{Fig:simulation}.  At constant voltage, the transformation time depends mainly on the thermodynamic driving force, or overpotential beyond critical voltage, and only weakly on the particle size. For small driving force, the particle initially lingers near the critical point. The transformation then accelerates exponentially due to the release of elastic energy (linearly growing overpotential from the tilted voltage plateau), but a detailed theory is beyond the scope of this paper. }

{\it Application to} \ce{LiFePO4}. --
\hl{Parameters for the phase-field model were estimated previously by fitting a regular solution model for $f(c)$ and a gradient energy to experimental solubility data {\cite{Cogswell2012}}. Based on the fitting, a chemical solubility limit of $X=.01$ and coherent solubility limit of $X=.09$ were estimated, in agreement with experiment. Using these phase-field parameters} and the surface energies in Table \ref{fig:wetting} (Eq.~(\ref{Eq:boundary_condition}) reduces to Dirichlet boundaries with $c_s=0$ or $c_s=1$ with the application of Eq. \ref{Eq:grad_c}), we consider realistic \ce{LiFePO4} particle geometries characterized by Smith \textit{et al.} \cite{Smith2012}. Simulation of intercalation dynamics is presented Fig. \ref{Fig:simulation} for particle shape C3, using the 2D depth-averaged model described in Methods \cite{Cogswell2012,Bazant2013}. The lithiated phase originates from the side facets, and intercalation waves ~\cite{singh2008,Bai2011} propagate inward, with phase boundary orientation tending to occur along elastically preferred $(101)$ planes.

Figure \ref{Fig:critical_potential}(a) shows calculated equilibrium discharge curve for \ce{LiFePO4} nanoparticles with circular cross-sections of varying diameter. The critical potential on discharging, or lithiation from a low state of charge, is reduced in smaller particles, as a result of lithium adsorption on the side facets, which competes with coherency strain to facilitate nucleation. (The same effect is not seen during charging, or delithiation from high concentration, since the 2D depth-averaged model does not model dewetting of the (010) surface.)

The size effect resulting from surface adsorption is calculated for a variety of particle shapes in Fig. \ref{Fig:critical_potential}(b).  Each \ce{LiFePO4} geometry is treated as a prism, with the lithiated side facets contributing to the wetted surface area $A$.  The calculated critical potentials are close to a master line given by Eq. \ref{Eq:critical_potential}, so the $A/V$ ratio is much more important than the precise shape. As shown in Methods, each surface layer is localized and planar, so its tension (energy/area) is largely unaffected by the bulk geometry. 

At a critical wetted area-to-volume ratio,  $A/V = L_c^{-1} \approx \unit[.18]{/nm}$, the critical potential vanishes, $\Delta\phi^*=0$ . Below the critical size $L_c$ (analogous to the critical radius in classical nucleation theory), the nanoparticle becomes dominated by its surface properties, rather than its bulk properties.  For prism-shaped \ce{LiFePO4} particles, this corresponds to mean particle diameter, $d_c = 4 L_c=\unit[22]{nm}$.  A delithiated particle below this size will spontaneously lithiate and remain lithiated at equilibrium. Interestingly, this is the same size at which phase separation is suppressed in \ce{LiFePO4} \cite{Wagemaker2011,Cogswell2012}. It is also roughly the cutoff size for anomalous phase behavior in bimetallic nanoparticles~ \cite{Dick2002,Shibata2002}.

The calculations in Fig. \ref{Fig:critical_potential}(b) are compared with extensive experimental data in Fig. \ref{Fig:critical_potential}(c) without adjusting any parameters.  For each experiment,  $A/V$ is found by approximating the particles as prisms and taking $A$ to be the area of the side facets. Since the calculation represents an ideal thermodynamic limit, experimental data points are expected to lie on or below the theoretical line (Eq. \ref{Eq:critical_potential}). 

The collapse of experimental data in Fig. \ref{Fig:critical_potential}(c) strongly supports the theory. No experimental points lie significantly above the simulated limit, and most lie very close to it. Some points also lie below, as  expected for two reasons. Firstly, it is likely that some experiments overestimate the size of the active particles. If the system has a distribution of particle sizes, a consequence of Eq. \ref{Eq:critical_potential} is that small particles will be transformed first at lower potentials. Therefore measuring an average particle size may not be sufficient to characterize a porous electrode. If measurements are made while only a fraction of the total capacity is cycled (as done by Dreyer \textit{et al.} \cite{Dreyer2010}), the properties of the smallest particles in the system may inadvertently be measured. Secondly, defects may play an important role by reducing coherency strain and decreasing the barrier for intercalation. Meethong \textit{et al.} measured different amount of retained strain in different samples \cite{Meethong2007,Meethong2007a}. If the amount of retained strain decreases, Eq. \ref{Eq:critical_potential} predicts the bulk phase transformation barrier $\Delta\phi_\infty^*$ will also decrease. In the limit of zero strain, the particle will transform at the miscibility gap where $\Delta\phi_\infty^*=0$.


\textit{Discussion} -- \ce{Li_XFePO4} was initially thought to be a slow-rate material due to  kinetic limitations of nucleation and growth \cite{Padhi1997}.  However, this hypothesis is inconsistent with the rapid rate improvements from smaller nanoparticles \cite{bruce2008}, doping to improve electrical conductivity \cite{Chung2002}, and the use of exotic surface coatings \cite{Kang2009, Sun2011}. These modifications are not expected to significantly alter growth kinetics.

Recently there has been a shift in thinking to the opposite extreme, with ultra-fast rate capabilities attributed to particles that never phase-separate. Malik and Ceder \cite{Malik2011} calculated a solid solution pathway using quantum Monte Carlo and argued against the possibility of nucleation and growth by applying CNT at the bulk spinodal.  A significant result of phase-field theory, however, is that the nucleation barrier goes to zero -- and CNT fails catastrophically -- at the spinodal point, in agreement with experiment \cite{Cahn1959}. The argument was also based on a phase boundary energy of $\gamma_{\alpha\beta}=\unit[960]{mJ/m^2}$ from {\it ab initio} calculations \cite{Wagemaker2009}, which is likely  more than an order of magnitude too large. We have inferred $\gamma_{\alpha\beta} = 39$ mJ/m$^2$ from experimental morphology and solubility data \cite{Cogswell2012}, consistent with the bound $\gamma_{\alpha\beta}< 200$ mJ/m$^2$ for coherent interfaces~\cite{Porter2009}.  (From a modeling perspective, this illustrates the limitations of \textit{ab initio} calculations that are restricted to a few unit cells of material, often at zero temperature, although our results also demonstrate the remarkable accuracy of the {\it ab initio} bulk and surface properties, used to parameterize our phase-field theory.) 

Dreyer \textit{et al.} \cite{Dreyer2010} also advocated for particles that never phase separate based on the observation of a \unit[20]{mV} ($\Delta\phi^*=\unit[10]{mV}$) ``thermodynamic'' hysteresis \ce{Li_XFePO4}.  Although the basic picture of mosaic phase separation may hold, we have shown that there is no unique value of the critical potential.  In order to be consistent with experimental observations of coherent phase separation~\cite{Cogswell2012}, the bulk critical potential must be much larger than \unit[10]{mV}, and the inferred value $\Delta\phi^*_\infty = 35$ mV  is confirmed here by an independent analysis of nucleation (Fig. \ref{Fig:critical_potential}).  Indeed,  the discrepancy over hysteresis can be resolved by combining our theory of nucleation with porous electrode theory based on non-equilibrium thermodynamics~\cite{ferguson2012},  as will be reported elsewhere. 

\textit {Conclusion } -- 
We have developed a quantitative phase-field theory of nucleation in \hl{single-crystal} nanoparticles.  A key observation is that complete ``wetting" by one solid phase is typically favored at each surface, so that nucleation corresponds to the instability of a wetted surface layer. In order to overcome coherency strain, the nucleation barrier becomes a linear function of the surface-to-volume ratio, which implies that nanoparticles tend to transform in order of increasing size. The theory is confirmed by collapsing disparate experimental data for nucleation in  \ce{Li_XFePO4} nanoparticles, without any adjustable parameters. Beyond important applications to Li-ion batteries, however, the theory also has broader relevance for nanotechnology. It provides the basic principles to design solid nanostructures with desired phase behavior under different operating conditions, by controlling the elastic and surface properties of the component materials. 

\begin{small}

\section{Methods}
{\it Dynamical Model. -- }
The mean intercalation rate  at the surface is related to the local surface concentration, stress state, and overpotential $\eta = \frac{\delta G}{e \rho_s \delta c}$ by the phase-field theory of electrochemical kinetics~\cite{Bazant2013}.  At equilibrium, $G$ is minimized, and $\eta=0$. To model intercalation dynamics (Fig. ~\ref{Fig:simulation}), we assume reaction limitation for anisotropic  \ce{LiFePO4} nanoparticles (with fast diffusion and no phase separation along the [010] crystal axis and negligible diffusion along the [100] and [001] axes) and solve the electrochemical Allen-Cahn reaction equation~\cite{Bazant2013,singh2008,Bai2011,Cogswell2012}, 
\begin{equation}
\frac{\partial c}{\partial t}=\frac{2}{e} J_0(c,\nabla^2 c,\vec{u}) \sinh\left(\frac{\delta G}{e \rho_s \delta c}\right) + \xi  \label{eq:EAC}
\end{equation}
for the depth-averaged concentration $c(x,y)$ over the active (010) facet. The right side of Eq. ~\ref{eq:EAC} is a generalized Butler-Volmer rate for symmetric electron transfer, where $J_0$ is the exchange current per area~\cite{Bazant2013}. Stochastic intercalation is modeled with Langevin noise $\xi$, which facilitates nucleation or spinodal decomposition~\cite{Cogswell2012,Bai2011}. The strain field is determined by mechanical equilibrium, $\frac{\delta G}{\delta \vec{u}}=\nabla\cdot\sigma=0$, and zero surface traction, $\hat{n}\cdot\sigma=0$, for a solid particle in a liquid electrolyte~\cite{Cogswell2012,Tang2011}.

{\it Derivation of the Critical Potential. -- }
For functionals that do not depend explicitly on $x$, the Beltrami Identity is an integrated form the the Euler equation that applies at equilibrium in 1D systems. Application of this relation to Eq. \ref{Eq:energy_functional} produces:
\begin{equation}
 \rho_s\left[f(c)+e\Delta\phi c\right]-\frac{1}{2}\kappa(\nabla c)^2+\frac{1}{2}C_{ijkl}\epsilon_{ij}\epsilon_{kl}=C
 \label{Eq:Euler_equation}
\end{equation}
where $C$ is a constant whose value can be determined by considering the boundary condition at $x=L$, which is stress-free and $c(L)=c_0$. Thus $C=\rho_s\left[f(c_0)+e\Delta\phi c_0\right]$, where $c_0$ is the bulk concentration in equilibrium with $\Delta\phi$.  Solving Eq. \ref{Eq:Euler_equation} for the gradient energy produces:
\begin{equation}
 \frac{1}{2}\kappa(\nabla c)^2=\rho_s\left[\Delta f+e\Delta\phi\Delta c\right]+\frac{1}{2}C_{ijkl}\epsilon_{ij}\epsilon_{kl}
 \label{Eq:gradient_energy}
\end{equation}
where $\Delta f=f(c)-f(c_0)$, and $\Delta c=c-c_0$. Solving Eq. \ref{Eq:gradient_energy} for $\nabla c$ produces:
\begin{equation}
 \nabla c=\pm\sqrt{\frac{2}{\kappa}\left(\rho_s\left[\Delta f+e\Delta\phi\Delta c\right]+\frac{1}{2}C_{ijkl}\epsilon_{ij}\epsilon_{kl}\right)}
 \label{Eq:grad_c}
\end{equation}

\hl{The excess free energy of the adsorption layer $\sigma$ {\cite{Cahn1977}} (which differs from the surface energy $\gamma$)} is the difference in energy between a homogeneous system and a system with the adsorption layer:
\begin{equation}
 \begin{split}
  \sigma=G-\int_V\rho_s\left[f(c_0)+e\Delta\phi c_0\right]\, dV
 \end{split}
 \label{Eq:surface_excess}
\end{equation}
Substitution of Eq. \ref{Eq:gradient_energy} to eliminate the gradient energy leads to:
\begin{equation}
 \sigma=\gamma (c_s)+2\int_0^L \rho_s\left[\Delta f+e\Delta\phi\Delta c\right]+\frac{1}{2}C_{ijkl}\epsilon_{ij}\epsilon_{kl}\, dx
 \label{Eq:surface_excess_function}
\end{equation}
\hl{which is minimized at equilibrium}. Eq. \ref{Eq:grad_c} can be used to change the variable of integration in Eq. \ref{Eq:surface_excess_function} from $x$ to $c$:
\begin{equation}
 \sigma=\gamma (c_s)+\sqrt{\kappa}\int_{c_s}^{c_0} \sqrt{2\rho_s\left[\Delta f+e\Delta\phi\Delta c\right]+C_{ijkl}\epsilon_{ij}\epsilon_{kl}}\, dc
 \label{Eq:surface_excess_c}
\end{equation}

For an adsorption layer at a surface in a semi-infinite system, the elastic energy can be approximated as $C_{ijkl}\epsilon_{ij}\epsilon_{kl}=B_0(c-c_0)^2$ \cite{Cahn1961}. The stability of the adsorption layer is related to the sign of $\frac{d\sigma}{dc}$, and $\sigma$ is maximized when $\frac{\delta\sigma}{\delta c}=0$:
\begin{equation}
 \frac{\delta\sigma}{\delta c}=2\left(\left[\rho_sf'(c)+e\Delta\phi\right]+B_0(c-c_0)\right)=0
 \label{Eq:dsigma_dc}
\end{equation}
The equilibrium condition $e\Delta\phi=-f'(c_0)$ can be applied in the bulk, far from the interface:
\begin{equation}
 \rho_sf'(c)+B_0c=\rho_sf'(c_0)+B_0c_0
 \label{Eq:coherent_common_tangent}
\end{equation}
which is the common tangent construction for coherent binary solids \cite{Cahn1961}.  Thus the adsorption layer will become unstable and grow at the coherent solubility limit (miscibility gap). The energy needed to reach the coherent solubility limit is the barrier energy to initiate transformation of the particle.

The dependence of the surface excess $\sigma$ on the size of the system can be determined by examining the sign of $\frac{\partial\sigma}{\partial L}$, which is found using the fundamental theorem of calculus:
\begin{equation}
 \frac{\partial\sigma}{\partial L}=2\rho_s\left[\Delta f+e\Delta\phi\Delta c\right]+C_{ijkl}\epsilon_{ij}\epsilon_{kl}\Big |_{x=L}
\end{equation}
Since the boundary condition at $x=L$ is stress-free and $c(L)=c_0$:
\begin{equation}
 \frac{\partial\sigma}{\partial L}=0
\end{equation}
Thus $\sigma$ is independent of the system size $L$.

From Eq. \ref{Eq:surface_excess}, we can see that:
\begin{equation}
 G=\int_V\rho_s\left[f(c_0)+e\Delta\phi c_0\right]\, dV+\sigma A
 \label{Eq:G}
\end{equation}
The terms inside the integral are constants, and the integral can be evaluated:
\begin{equation}
 G=\rho_s\left[f(c_0)+e\Delta\phi c_0\right]V+\sigma A
\end{equation}

Now set $\frac{dG}{dc_0}=0$ for $c_0=c_0^*$ and solve for $\Delta\phi^*$:
\begin{equation}
 \frac{dG}{dc_0}=\rho_s[f'(c_0^*)+e\Delta\phi]V+\frac{d\sigma}{dc_0}(c_0^*)A=0
\end{equation}
\begin{equation}
 e\Delta\phi^*=-f'(c_0*)+\frac{1}{\rho_s}\left(\frac{d\sigma}{dc_0}(c_0^*)\right)\frac{A}{V}
\end{equation}
where $c_0^*$ is the concentration at the coherent solubility limit, determined by the coherent common tangent construction (Eq. \ref{Eq:coherent_common_tangent}).

Using Eq. \ref{Eq:surface_excess_c},  an approximation for elastic energy $C_{ijkl}\epsilon_{ij}\epsilon_{kl}=B_0(c-\bar{c})^2$ \cite{Cahn1961}, and applying the fundamental theory of calculus, an estimate for $\frac{d\sigma}{dc_0}$ can be obtained:
\begin{equation}
 \frac{d\sigma}{dc_0}(c_0^*)=\sqrt{\kappa B_0(c_0^*-\bar{c})}
\end{equation}
where $\bar{c}=\frac{1}{V}\int_Vc\,dV$ is the mean value of $c$.

\section{Acknowledgements}
\begin{acknowledgments}
We are grateful to Kyle Smith for providing advice on \ce{LiFePO4} particle geometries and for kindly sharing his Matlab code with us.  This work was supported by the National Science Foundation under Contracts DMS-0842504 and DMS-0948071 and by a seed grant from the MIT Energy Initiative.
\end{acknowledgments}

\end{small}

\bibliography{PRL_LFP_nucleation}
\end{document}